\documentclass[11pt]{article} 

\usepackage{geometry} 
\usepackage{amsmath} 
\usepackage{amsthm} 
\usepackage{amssymb} 
\usepackage{slashed}
\usepackage{dsfont}
\usepackage{bbold}
\usepackage[super]{nth}
\usepackage{blindtext}

\author{Thomas E. Williams\thanks{t.e.williams@doctoral.uj.edu.pl} \\ Institute of Physics, Jagiellonian University, \\
prof. Stanis\l awa \L ojasiewicza 11, 30-348 Krak\'ow, Poland}
\title{A superspace formulation of SUSY in NCG with spectral action}

\theoremstyle{definition}

\theoremstyle{definition}

\theoremstyle{definition}

\theoremstyle{definition}

\newcommand{\Tr}{\textsf{Tr}}
\newcommand{\e}{\mathrm{e}}

\begin{document}
\maketitle  

\section{Introduction}

Noncommutative geometry (NCG) is a relatively new branch of mathematics, the bulk of which was worked out in the latter half of the previous century by French mathematician Alain Connes\cite{Connes1}.
In its essence, noncommutative geometry tells us how to abstract all pertinent information about a classical differentiable manifold to the level of operators and abstract algebras of functions defined over its coordinates.  Furthermore, Connes goes on to formulate a set of conditions for which the converse is also achieved.  Via his celebrated \textit{reconstruction theorem}, one may recover the geometric information about the underlying manifold from a set of purely algebraically defined quantities.

In the formalisms of operator theory and abstract algebra there is no need to restrict to the study of commutative algebras only, thus by considering noncommutative algebras, a broader class of ``geometric spaces" is studied, for which a classical geometric description is impossible.  Such spaces are frequently referred to as \textit{noncommutative geometries} or \textit{noncommutative manifolds}.  As it turns out, it is in this class of noncommutative spaces (or more correctly a restricted subclass known as \textit{almost-commutative manifolds} or \textit{AC-manifolds}) in which Connes found a unified theory of the complete Standard Model (SM) of particle physics coupled with classical (Einsteinian) gravity \cite{SMneutrinos}.

In the present paper the ideas and methods of NCG  are combined and applied to construct a toy model possessing an interesting (even if not thus far experimentally observed) virtue -- a supersymmetry.

Simply stated, supersymmetry (SUSY) is a proposed symmetry of nature which relates to each boson (a particle with integer spin) an associated partner particle with half integer spin (a fermion), and vice versa.  Although initially posited as a meson/baryon symmetry in the theory of hadrons \cite{Miyazawa}, it was reincarnated several years later as a global spacetime symmetry in the context of quantum field theories (QFTs).  It is perhaps in the work of Julius Wess and Bruno Zumino \cite{WZ}, that supersymmetry really came of age.  Their work provided the first example of a four dimensional, supersymmetric quantum field theory with interactions.

While the defining principle of supersymmetry is concisely stated, its simplicity is disproportionate to its value.  With the additional principle of supersymmetry, many of the curiosities and apparent inconsistencies of the SM are readily explained. The unexpectedly low mass of the Higgs particle, the hierarchy problem, and the nonunification of the gauge coupling constants at high energies to name a few. Additionally, SUSY provides a candidate for the particle(s) responsible for dark matter, the only possible  ``workaround" for the Coleman-Madula theorem, and a ray of hope for (potentially) physically relevant string theories which without SUSY would be out of business \cite{Primer,1001Lessons}.

With the exception of \cite{PTEP}, the work by Wim Beenakker, Thijs van den Broek, and Walter D. van Sujlekom, \textit{Supersymmetry and Noncommutative Geometry} \cite{Beenekker}, seems to be the only previous attempt to reconcile NCG with SUSY which makes use of the \textit{spectral action}.  Starting from the requirement that the resulting spectral action functional be supersymmetric, Beeneker et al. provide a classification of all supersymmetric AC-geometries whose particle content this ensures.  Meanwhile avoiding mention of \textit{superfields} or \textit{supermanifolds}.

We concur with their assessment that due to its vital importance to the predictive power of the noncommutative method and its proven success in producing the SM, the spectral action deserves a distinguished place at the table.  However, we are also of the opinion that a \textit{superspace} should be the natural starting point for combining SUSY with NCG since it elegently encodes the SUSY transformation as geometric translations of its coordinates \cite{Rogers,DeWitt}.

Several attempts have been made to combine SUSY with NCG which do relax the definition of AC-manifold sufficiently to allow the algebra of superfields over a supermanifold as the primary object of study, but with the exception of the afore mentioned PTEP article, they fail to employ the spectral action \cite{SUSYNCG1, SUSYNCG2,SUSYNCG3,SUSYNCG4}.  In that we start from an algebra of superfields defined on superspace coordinates, and ultimately require a supersymmetric spectral action functional, we feel that our work is distinguished from previous approaches.

Herein, a framework is proposed for incorporating a superspace formulation of the principle of SUSY into the formalism of NCG with a spectral action.  An example of this framework is explored wherein the spaces considered are kept simple in order to highlight the main ideas and methods employed.  In future work, spaces with more complicated structure will be explored with an eye toward theories of potentially physical relevance.

\section{3d superfields and the fermionic action}

  \subsection{Essential elements of NCG}
Due to the enormity and mathematical depth of the subject, the reader is referred to the existing literature for a more comprehensive introduction \cite{Elements,WvS} and instead give herein only a rapid overview of the elements which will be necessary for what follows.

As previously stated, the recipe followed will be the \textit{AC-geometry} approach with conditions suitably relaxed to accommodate \textit{superalgebras} over superspace coordinates.  With this in mind, recall that the AC-geometry approach begins with a \textit{total space} of the form
\[
M\times F,
\]
where $M$ is a compact Reimannian spin manifold and $F$ is some finite, discrete topological space.

The next step is to pass to a set of algebraic structures which equivalently describes this space, known as a \textit{spectral triple}. This is done in 2 steps. First, to $M$ and $F$, associate the spectral triples
\[
\mathcal{M}\equiv(\mathcal{A}_M,\mathcal{H}_M,\mathcal{D}_M) \quad \text{and} \quad \mathcal{F}\equiv(\mathcal{A}_F,\mathcal{H}_F,\mathcal{D}_F),
\]
where $\mathcal{M}$ consists of a unital, associative $*$-(super)algebra, $\mathcal{A}_M$, faithfully represented on a Hilbert space of bounded operators, $\mathcal{H}_M$,  and a self adjoint (Hermetian) operator, $\mathcal{D}_M:\mathcal{H}_M\to \mathcal{H}_M$, often taken to be a \textit{Dirac operator}, with compact resolvent, and such that  $[\mathcal{D}_M,a]$ is bounded for any $a\in\mathcal{A}_M$, and where $\mathcal{F}$ consists of a finite dimensional, unital, associative $*$-(super)algebra, $\mathcal{A}_F$, faithfully represented on a finite dimesional Hilbert space, $\mathcal{H}_F$, upon which acts a symmetric operator $\mathcal{D}_F$.  Then the spectral triple encoding the structure of the total space is given by the tensor product
\[
\mathcal{M}\otimes\mathcal{F}\equiv(\mathcal{A}, \mathcal{H}, \mathcal{D})\equiv(\mathcal{A}_M\otimes\mathcal{A_F}, \mathcal{H}_M\otimes\mathcal{H}_F, \mathcal{D}_M\otimes\mathcal{D}_F),
\]
where $\mathcal{D}\equiv\mathcal{D}_M\otimes\mathcal{D}_F\equiv\mathcal{D}_M\otimes \hbox{\boldmath $1$}_{ F}+\gamma_M\otimes \mathcal{D}_F$ is called the Dirac operator of the AC-manifold.

A spectral triple is said to be \textit{even} if the Hilbert space is equipped with a $\mathbb{Z}_2$-grading (an operator $\gamma:\mathcal{H}\to\mathcal{H}$, such that $\gamma^2=1$ ) which satisfies $[\gamma,a]=0$ and $\{\gamma,\mathcal{D}\}=0$.  Also, a spectral triple is called \textit{real}, if the Hilbert space admits a \textit{real structure}, that is, an anti-unitary operator, $J:\mathcal{H}\to\mathcal{H}$ such that $J^2=\epsilon$, $J\mathcal{D}=\epsilon'\mathcal{D}F$, and in the case that the spectral triple is even, $J\gamma=\epsilon''\gamma J$, where $\epsilon$, $\epsilon'$, and $\epsilon''$ are each $\pm 1$ and together determine the \textit{KO-dimension} of the spectral triple. Moreover, it is required that the \textit{commutant property} (or \nth{0}-order condition) and the \textit{\nth{1}-order condition} hold,
\[
[a,b^\circ]=0, \quad \text{and} \quad [[\mathcal{D},a],b^\circ]=0,
\]
where $a,b\in\mathcal{A}$, and $b^\circ\equiv Jb^*J^{-1}$ implements a right action of $\mathcal{A}$ on $\mathcal{H}$.

Finally then, the object of central importance to this story, namely a real, even spectral triple
\[
(\mathcal{A},\mathcal{H},\mathcal{D};\gamma,J),
\]
where $\gamma\equiv\gamma_M\otimes\gamma_F$, and $J\equiv J_M\otimes J_F$, may be written down (generally).

In noncommutative geometry, the \textit{gauge fields} arise by considering \textit{Morita (self-)equivalence} of the algebra, meanwhile the \textit{gauge group} implements unitary equivalence of spectral triples (which is itself an instantiation of Morita equivalence). Briefly, the above real, even spectral triple is equivalent, up to Morita self-equivalence, to
\[
(\mathcal{A},\mathcal{H},\mathcal{D}_A;\gamma,J),
\]
where $\mathcal{D}_A=\mathcal{D}+A+\epsilon'J A J^{-1}$ is the \textit{inner fluctuated} Dirac operator and $A\in \Omega^1_\mathcal{D}(\mathcal{A})\equiv\{\sum_i a_i[\mathcal{D},b_i] : a_i,b_i \in \mathcal{A} \}$ are the \textit{gauge fields}, or \textit{inner fluctuations} of the Dirac operator, $\mathcal{D}$.  Meanwhile,
\[
(\mathcal{A},\mathcal{H},U\mathcal{D}U^*;\gamma,J),
\]
is a unitarily equivalent triple obtained by an element of the gauge group $U=uJuJ^{-1}$ where $u$ is a unitary element of $\mathcal{A}$.  Ultimately, $U\mathcal{D}U^*=\mathcal{D}_A$ for $A=u[\mathcal{D},u^*]$.

Physics then emerges from the noncommutative formalism in the form of a \textit{Lagrangian} constructed from the \textit{action functional},
\[
S=S_b+S_f\equiv\Tr \left( f\left( \frac{\mathcal{D}_A}{\Lambda}\right)\right)+\left\langle\xi,\mathcal{D}_A \xi\right\rangle,
\]
where $\xi\in\mathcal{H}$ and $f$ is some sufficiently well behaved function. This action consists of the \textit{spectral action}, $S_b$, responsible for the bosonic terms, and the \textit{fermionic action}, $S_f$, taking care of fermionic particle content. The former is usually evaluated by heat kernal methods and is spectral in the sense that it counts eigenvalues of the fluctuated Dirac operator up to some predetermined cut-off, $\Lambda$.

  \subsection{The $\Lambda_\infty$ algebra}
The \textit{Grassmann algebra}, $\Lambda_\infty(\mathbb{C})$, (hereafter abbreviated as $\Lambda_\infty$), is the unital, associative algebra generated by a countably infinite set of anti-commuting variables $\xi^i,$ that is,
\[ \xi^i \xi^j+\xi^j \xi^i =0, \quad \text{for all} \quad i,j \in \mathbb{N}.
\]

Each element $g\in \Lambda_\infty$ may be written as the sum of its \textit{body} and \textit{soul}, $g=g_B+g_S \in \Lambda_\infty^B \oplus \Lambda_\infty^S$, where
\[
g_S=\sum_{k=1}^\infty \frac{1}{k!} c_{i_1 i_2 \dots i_k} \xi^{i_1} \xi^{i_2} \dots \xi^{i_k}, \quad \text{and} \quad g_B, \: c_{i_1 i_2 \dots i_k} \in \mathbb{C.}
\]
Alternatively, $\Lambda_\infty$ may be decomposed into the direct sum of an \textit{even} subalgebra and an \textit{odd} subset, $\Lambda_\infty=\Lambda_\infty^e\oplus \Lambda_\infty^o$, where $\Lambda_\infty^e$ consists of $\Lambda_\infty^B$ and elements of $\Lambda_\infty^S$ with an even number of generating elements, $\xi ^i$, and likewise $\Lambda_\infty^o$ consists of elements of $\Lambda_\infty^S$ with an odd number of generating elements.

There are several possible involutive maps on $\Lambda_\infty$ which make it a $*$\textit{-algebra}, i.e. For any $g,h \in \Lambda_\infty$, $(gh)^*=h^*g^*$ and $(g^*)^*=g$. For now, define $*:\Lambda_\infty\to \Lambda_\infty$ to be $g\mapsto g^*=g_B^*+g_S^*$, where $g_B^*$ is ordinary complex conjugation, and
\[
g_S^*=\sum_{k=1}^\infty \frac{1}{k!} c_{i_1 i_2 \dots i_k}^* \xi^{i_k} \xi^{i_{k-1}} \dots \xi^{i_1}= \sum_{k=1}^\infty \frac{(-1)^{\frac{k(k-1)}{2}}}{k!} c_{i_1 i_2 \dots i_k}^* \xi^{i_1} \xi^{i_2} \dots \xi^{i_k}.
\]

The group of unitary elements of the Grassmann algebra is $\mathcal{U}(\Lambda_\infty)=\{ u\in \Lambda_\infty \:|\: uu^*=u^*u=1 \}, $
and since for $u\in \mathcal{U}(\Lambda_\infty)$, $u_B \neq 0,$ Grassmann unitaries are \textit{logarithmic}, that is, they are expressible as $u=\e^{ig}$ for some
\textit{real Grassmann number} $g,$ i.e.\ $g \in \Lambda_\infty$ satisfying  $g = g^*$.

\subsection{3d Minkowski spacetime}
In order to present the following ideas in a simple setting we choose to work
in a three-dimensional space with metric signature $(p,q)=(1,2)$, e.g. $\eta=\text{diag}(1,-1,-1)$. In this case
the universal cover of the Lorentz group is the group SL$(2,\mathbb{R})$ and the Dirac matrices
(i.e.\ generators of a matrix representation of the even graded \textit{Clifford algebra} $\text{Cl}^e_{1,2}(\mathbb{R})$) may be expressed via Pauli matrices, as
\[
\gamma^0 =\sigma^2, \gamma^1 = i\sigma^3, \gamma^2 = i\sigma^1.
\]
The spin representation of Lorentz transformations,  (i.e. $L$ such that $L^\text{T}\eta L=\eta$), is constructed in the standard way
\[
S(L) = \exp{\left\lbrace\frac{1}{4} \sum_{a<b} \xi_{ab}[\gamma^a,\gamma^b]\right\rbrace }, \quad \text{where} \quad \xi_{ab} = - \xi_{ba},
\]
and satisfies
\[
S^{-1}(L)\gamma^{m'}S(L) = L^{m'}_{\,\,\,m}\gamma^m,
\]
which guarantees covariance with the \textit{spinorial derivative} $D\equiv i\gamma^m\partial_m$.  Explicitly, $D$ acts on \textit{spinors} as $D\psi^\alpha(x)=i(\gamma^m)^\alpha_{\,\,\,\beta}\partial_m\psi^\beta(x)$, which transform under Lorentz transformations as $\psi^{\alpha'}(x)=S(L)^{\alpha'}_{\,\,\,\alpha}\psi^\alpha(L^{-1}x)$, so covariance means $D'\psi'(x')=S(L)D\psi(x)$.

For a Lorentz invariant, hermitian inner product, take $(\xi,\psi)\equiv i\bar{\xi}\psi$, where $\bar{\xi}\equiv\xi^\dagger\gamma^0$. $D$ is hermitian with respect to this product, i.e. $(D\chi,\psi)=(\chi,D\psi)$, and moreover, the complex conjugation operator, $C$, which acts by $C\psi=\psi^*$ is an anti-unitary operator, i.e. $(C\chi,C\psi)=(\chi,\psi)^*$.

  \subsection{The superspace, $\mathbb{R}^{3|2}$}
$\mathbb{R}^{3|2}$ is a coordinate space described by three commuting (bosonic) coordinates, say $x^1,x^2,x^3$, and two anticommuting (fermionic) coordinates, say $\theta^1$ and $\theta^2$
which are assumed to be independent of the variables $\xi^i$ generating the $\Lambda_\infty$ algebra, and to form a spinor of the 3d Lorentz group.
In the superspace construction, the global SUSY transformation correspond to translations of the superspace coordinates of the form
\begin{equation}
\label{SUSY:on:coordinates}
\delta\theta^\alpha\equiv\epsilon^\alpha \quad \text{and} \quad \delta x^m\equiv\theta_\alpha (\gamma^m)^\alpha_{\:\:\:\beta}\epsilon^\beta,
\end{equation}
in accordance with the transformation properties of $\theta^\alpha$ and $x^m$ under Lorentz transformations.
Component fields of a \textit{supermultiplet} (an \textit{irreducible representation} of the \textit{supersymmetry algebra}) are combined into a function of the superspace coordinates called a \textit{superfield},
\[
S(x,\theta) =f(x)+g_{\beta}(x)\theta^\beta+h(x)\theta\theta,
\]
for some component functions $f, g, h \in \Lambda_\infty$.  Here the convention adopted is
\[
\theta\theta \equiv \theta^2\theta^1 = \frac{1}{2}\varepsilon_{\alpha\beta}\theta^\beta\theta^\alpha =\frac{1}{2}\theta_\alpha\theta^\alpha, \quad \text{where} \quad \varepsilon_{12}=-\varepsilon_{21}=\varepsilon^{12}=-\varepsilon^{21}=1.
\]
Comparing the two expressions for the infintesimal SUSY variation of $S:$
\[
\delta S(x,\theta) = \delta f(x)+ \delta g_{\beta}(x)\theta^\beta+ \delta h(x)\theta\theta
\]
and
\[
\delta S(x,\theta) = S(x +\delta x, \theta + \delta\theta) - S(x,\theta),
\]
it is seen, in particular, that
\begin{equation}
\label{SUSY:highest:component}
\delta h(x) = \partial_m \left(g_\beta(\gamma^m\epsilon)^\beta\right).
\end{equation}
The fact that the $\theta\theta$ component of a superfield transforms under the SUSY variation as a total derivative implies that the integral,
\[
\int h\, d^3x \equiv \int S_{\theta\theta}\, d^3x,
\]
is invariant under supersymmetry transformation.

The infinitesimal SUSY variation of a superfield can be expressed through the first order differential operator $Q_\beta.$ Indeed, writing \
\begin{equation}
\delta S = [\epsilon^\beta Q_\beta, S],
\end{equation}
implies
\begin{equation}
Q_\beta=-\partial_\beta+(\theta\gamma^m)_\beta\partial_m.
\end{equation}
It is also useful to define the \textit{superspace covariant} derivative (or \textit{super-covariant} derivative),
\begin{equation}
D_\alpha = \partial_\alpha +(\theta\gamma^m)_\alpha\partial_m,
\end{equation}
which anticommutes with the generator of the SUSY transformation,
\begin{equation}
\label{anticommutator:Q:D}
\{D_\alpha,Q_\beta\} \equiv D_\alpha Q_\beta + D_\beta Q_\alpha = 0,
\end{equation}
and squares to a proportion of the spinorial derivative,
\[
\{D^\alpha,D_\beta\} = -2 (\gamma^m)^\alpha{}_\beta \partial_m.
\]

The present construction will utilize a \textit{spinor superfield},
\[
\Psi^\alpha (x,\theta)=\psi^\alpha (x) +F^\alpha_{\:\:\:\beta}(x)\theta^\beta +\chi^\alpha(x)\theta\theta,
\]
which under Lorentz transformations changes as $\Psi'(x,\theta)=S(L)\Psi(L^{-1}x,S(L)^{-1}\theta).$ The infinitesimal SUSY variation of the component fields of $\Psi(x,\theta)$ read
\begin{align*}
\delta\psi^\alpha              =& F^\alpha_{\:\:\:\beta}\epsilon^\beta, \\
\delta F^\alpha_{\:\:\:\beta}	  =&  \partial_m\psi^\alpha(\gamma^m\epsilon)^\rho\varepsilon_{\rho\beta} -\chi^\alpha\epsilon_\beta, \\
\text{and} \quad \delta\chi^\alpha              =& \partial_m F^\alpha_{\:\:\:\beta}(\gamma^m\epsilon)^\beta.
\end{align*}
Note that in order to have $\Psi(x,\theta)\in\Lambda_\infty^o$, it must be that $\psi^\alpha(x), \chi^\alpha(x) \in\Lambda_\infty^o$ and $F^\alpha_{\;\;\;\beta}(x)\in\Lambda_\infty^e$.

The components of the spinor superfield do not form an irreducible representation of the supersymmetry transformation.
They may be constrained by requiring $D_\alpha\Psi^\alpha = 0.$ This is compatible with the SUSY transformation of $\Psi$ since,
thanks to (\ref{anticommutator:Q:D}),
\[
D_\alpha\Psi^\alpha = 0 \hskip 5mm \Rightarrow \hskip 5mm  D_\alpha\delta\Psi^\alpha
=
D_\alpha\epsilon_\beta Q^\beta \Psi^\alpha =\epsilon_\beta Q^\beta D_\alpha\Psi^\alpha = 0.
\]
Explicitly,
\begin{align*}
D_\alpha \Psi^\alpha &= \partial_\alpha \Psi^\alpha + \theta_\beta(\gamma^m)^\beta_{\;\;\;\alpha}\partial_m \Psi^\alpha \\
&=F^\alpha_{\;\;\;\alpha} - \theta_\beta\chi^\beta + \theta_\beta(\gamma^m )^\beta_{\;\;\;\alpha}\partial_m \Psi^\alpha + (\gamma^m)^\beta_{\;\;\;\alpha}\partial_m F^\alpha_{\;\;\;\beta}\theta\theta,
\end{align*}
so that the components of the \textit{chiral spinor superfields}, ${\tilde\Psi}^\alpha,$ defined by the relation $D_\alpha{\tilde\Psi}^\alpha = 0,$ satisfy
\begin{align*}
&\Tr F \equiv F^\alpha_{\:\:\:\alpha}=0,  \\
&\chi^\beta=(\gamma^m )^\beta_{\:\:\:\alpha}\partial_m\psi^\alpha, \\
\text{and} \quad &\Tr \left[ \gamma^a \partial_a F\right] \equiv (\gamma^m )^\beta_{\:\:\:\alpha}\partial_m F^\alpha_{\:\:\:\beta}=0,
\end{align*}
and transform according to the rule
\begin{align*}
\delta\psi^\alpha              =& F^\alpha_{\:\:\:\beta}\epsilon^\beta, \\
\text{and} \quad \delta F^\alpha_{\:\:\:\beta}	  =&  \partial_m\psi^\alpha(\gamma^m\epsilon)^\rho\varepsilon_{\rho\beta} -(\gamma^m)^\alpha_{\:\:\:\rho}\partial_m \psi^\rho(x)\epsilon_\beta.
\end{align*}

  \subsection{The Dirac operator and the chiral restricted fermionic action}
For now, the Dirac operator will be the usual spinorial derivative on 3d Minkowski spacetime,
\[
\mathcal{D}_M\equiv D=i\gamma^m\partial_m,
\]
and it will act on the Hilbert space of chiral spinor superfields $\tilde{\Psi}(x,\theta)$ over $\mathbb{R}^{3|2}$. The chiral restricted fermionic action is then taken to be
\[
\left\langle \tilde{\Psi}, \mathcal{D}_M \tilde{\Psi} \right\rangle\equiv\left(\tilde{\Psi},\mathcal{D}_M \tilde{\Psi}  \right).
\]
The term of highest order in the Grassmann variables is by construction invariant under a supersymmetry transformation, and  it is calculated to be
\[
\left\langle \tilde{\Psi}, \mathcal{D}_M \tilde{\Psi} \right\rangle_{\theta\theta}= \left\langle \psi,i\gamma^m\partial_m\chi\right\rangle + \left\langle F_2+F_1, i\gamma^m\partial_m(F_1-F_2) \right\rangle + \left\langle \chi,i\gamma^m\partial_m\psi \right\rangle.
\]

\section{Inner fluctuations and the spectral action}

  \subsection{N-point superspace and the distance function}
Take the (unital, associative) $*$-algebra $\Lambda(F)$ of Grassmann number ($\Lambda_\infty $)-valued functions over a finite topological space $F$ consisting of $N$ distinct points and endowed with the discrete topology.  Let this algebra be equipped with pointwise linear multiplication, addition, and with involution as previously discussed, i.e. for any $f,g\in \Lambda(F)$ and $\lambda\in \mathbb{C}$,

\begin{itemize}
\item $(f+g)(x)=f(x)+g(x)$,
\item $(\lambda f)(x) = \lambda f(x)$,
\item $(fg)(x)=f(x)g(x)$.
\end{itemize}
Notice that for the case of a finite discrete space $F$, the map
\[
\Lambda(F)\ni \varphi \mapsto (\varphi(1),\varphi(2),\dots,\varphi(N))\in \Lambda^N\equiv\Lambda\oplus \Lambda\oplus \cdots\text{N-copies}\cdots \oplus \Lambda,
\]
is a $*$-algebra isomorphism, $\Lambda(F)\simeq \Lambda^N$.  The above copies of the Grassmann algebra may conveniently arranged as entries along the main diagonal of an $N\times N$ matrix

\[
\left( \begin{array}{cccc}
\varphi(1) & 0 & \cdots & 0 \\
0 & \varphi(2) & \cdots & 0 \\
\vdots & \vdots & \ddots & \vdots \\
0 & 0 & \cdots & \varphi(N)
\end{array} \right),
\]
so that pointwise multiplication and addition are simply matrix multiplication and addition, respectively.

Now, if $F$ is endowed with a metric $d_{ij}$, then there exists a representation, $\pi$ of $\Lambda(F)$ on a finite dimensional Hilbert space, and a bounded symmetric operator, $\mathsf{D}$, such that
\[
d_{ij}=\sup_{f\in\Lambda^N}\left\lbrace |f(i)-f(j)| : ||[\mathsf{D},\pi(f)]||\leq 1 \right\rbrace.
\]
This claim follows from the equality
\[
|| [D,\pi (f)]|| = \max_{k\neq l}\left\lbrace \frac{1}{d_{kl}}|\phi(k)-\phi(l)|\right\rbrace,
\]
which is proved by an induction argument following \cite{WvS} Thm 2.18. pp 19-20. Therefore, it also makes sense in the present context to speak of the Dirac operator as a fundamental object which determines the geometry of a (super)space.

Henceforth, $F$ is taken to be a $2$-point discrete space.

  \subsection{The total space spectral triple}
All the ingredients are now available to construct the spectral triple for the total space which is to be presently considered.  The base space spectral triple is characterized by the algebraic data
\[
\mathcal{M}^{3|2}\equiv\left( \mathcal{A}_M = \Lambda_\infty^e, \mathcal{H}_F, \mathcal{D}_M=i\gamma^m\partial_m;  \gamma_M\equiv\begin{pmatrix}1&0\\0&-1 \end{pmatrix}, J_M\equiv\begin{pmatrix}G&0\\0&G\end{pmatrix} \right),
\]
where $\mathcal{H}_M$ is the (Hilbert) space of spinor superfields, $\Psi(x,\theta)$, and where $G$ denotes Grassmann conjugation. And the finite space spectral triple is
\[
\mathcal{F}_F\equiv\left((\Lambda_\infty^e)^2, (\Lambda_\infty^o)^2, \mathcal{D}_F=0, \gamma_F = \begin{pmatrix}
1 &0\\
0&-1
\end{pmatrix}, J_F=\begin{pmatrix}
0 & G \\
G & 0
\end{pmatrix}\right),
\]
where the form of $J_F$ is for the case of KO-dimension 6, and $\mathcal{D}_F=0$ follows from the spectral triple for a 2-point finite space with 2 dimensional Hilbert space representation being equipped with a real structure.
The resulting triple for the total space is then
\[
\mathcal{M}^{3|2}\otimes \mathcal{F}_F\equiv\left(\mathcal{A}=(\Lambda_\infty^e)^2, \mathcal{H}=(\Psi(x,\theta))^2, \mathcal{D}=\mathcal{D}_M\otimes\hbox{\boldmath $1$}_{ F}, \gamma=\gamma_{M}\otimes\gamma_F, J=J_M\otimes J_F \right).
\]
It should be stressed that the tensor product used here is over Grassmann numbers rather than the usual one over $\mathbb{C}$.

  \subsection{Fluctuating the Dirac operator}
The Dirac operator for the total space,  $\mathcal{D}=\mathcal{D}_M\otimes  \hbox{\boldmath $1$}_{ F}$, where $\mathcal{D}_M=i\gamma^m\partial_m$, may be written in a matrix form for a 2 point finite space geometry, as
\[ \mathcal{D} = i\gamma^m \begin{pmatrix}
\partial_m & 0 \\ 0 & \partial_m
\end{pmatrix}.
\]
To calculate $\mathcal{D}_A= \mathcal{D} + A + JAJ^{-1}$ the form of $A \in \Omega^1_\mathcal{D} (\mathcal{A})\equiv\{ a[\mathcal{D},b] : a,b \in \mathcal{A}=(\Lambda^e_\infty)^2 \}$ is needed.
So, taking
\[
a=\begin{pmatrix}
a_1 & 0 \\
0 & a_2
\end{pmatrix} \quad \text{and} \quad
b=\begin{pmatrix}
b_1 & 0 \\
0 & b_2
\end{pmatrix},
\]
gives
\[
A=a[\mathcal{D},b]=\begin{pmatrix}
ia_1\gamma^m \partial_m b_1 &0 \\
0 & ia_2\gamma^m \partial_m b_2
\end{pmatrix}.
\]
Now, in the case of a Grassmann number valued algebra over a 2 point finite space geometry, the restrictions $u_1u_2^*=u_2u_1^*=-1$ and $u_1u_1^*=u_2u_2^*=-1$ which arise from the condition $\epsilon=-1\implies J_F^2=-\hbox{\boldmath $1$}_{ F}$ in KO-dimensions 2, and 4 respectively, cannot be satisfied except trivially, and thus are excluded in the present situation.

However, using the form of $J_F$ for KO-dimension 6,
\[
JAJ^{-1}=\begin{pmatrix}
ia_2^*\gamma^m \partial_m b_2^* & 0 \\
0 & ia_1^*\gamma^m \partial_m b_1^*
\end{pmatrix}.
\]
Then, the fact that $A+JAJ^{-1}$ is traceless, (this follows from the hermiticity of $A$), implies that
\[
(a_i \partial_m b_i)^* = -a_i \partial_m b_i,
\] and so,
\[
A+JAJ^{-1} = \begin{pmatrix}
i\gamma^m (a_1\partial_m b_1 - a_2\partial_m b_2) & 0 \\
0 & -i\gamma^m (a_1\partial_m b_1 - a_2\partial_m b_2)
\end{pmatrix}.
\]
Finally then, the fluctuated Dirac operator for KO-dimension 6 is
\begin{align*}
\mathcal{D}_A &=i\gamma^m \begin{pmatrix}
\partial_m & 0 \\ 0 & \partial_m
\end{pmatrix}+\begin{pmatrix}
\gamma^m A_m & 0 \\
0 & -\gamma^m A_m
\end{pmatrix} \\
&=\mathcal{D}+\gamma^m A_m \otimes \gamma_F \quad \text{where} \quad A_m =i(a_1\partial_m b_1 - a_2\partial_m b_2).
\end{align*}
Similarly,
\[
JAJ^{-1}=\begin{pmatrix}
ia_1^*\gamma^m \partial_m b_1^* & 0 \\
0 & ia_2^*\gamma^m \partial_m b_2^*
\end{pmatrix},
\]
is obtained by using the form of $J_F$ for KO-dimension 0.
This time, the previously invoked trace free condition results in a trivial fluctuation, i.e. $\mathcal{D}_A=\mathcal{D}$.  Instead, let $a_1\slashed{D}b_1=-a_2\slashed{D}b_2$ and $(a_1\slashed{D}b_1)^*=-(a_1\slashed{D}b_1)^*$ so that $A+JAJ^{-1}$ is again traceless as required.
Changing labels so that $a_2=-a_1$ and $b_2=b_1$,
\[
A+JAJ^{-1} = \begin{pmatrix}
i\gamma^m (a_1\partial_m b_1 - a_1^*\partial_m b_1^*) & 0 \\
0 & -i\gamma^m (a_1\partial_m b_1 - a_1^*\partial_m b_1^*)
\end{pmatrix}.
\]
As before then, the fluctuated Dirac operator for KO-dimension 0 may be written as
\begin{align*}
\mathcal{D}_A &=i\gamma^m \begin{pmatrix}
\partial_m & 0 \\ 0 & \partial_m
\end{pmatrix}+\begin{pmatrix}
\gamma^m A_m & 0 \\
0 & -\gamma^m A_m
\end{pmatrix} \\
&=\mathcal{D}+\gamma^m A_m \otimes \gamma_F \quad \text{where this time} \quad A_m =i(a_1\partial_m b_1 - a_1^*\partial_m b_1^*).
\end{align*}

  \subsection{The gauge group and chiral superfield covariance}
Considering the finite space, $\mathcal{F}_F $, associated with the $2$-point discrete topological space $F$, take $\mathcal{U}(\mathcal{A}_F)$ to be the unitary elements of $\mathcal{A}_F$, i.e. $u\in\mathcal{U}(\mathcal{A}_F)$ which have the form
\[u=\begin{pmatrix}
u_1 & 0 \\
0 & u_2
\end{pmatrix} = \begin{pmatrix}
\e^{ig^{(1)}} & 0 \\
0 & \e^{ig^{(2)}}
\end{pmatrix},
\]
where $g^{(1)}$ and $g^{(2)}$ are real, even, Grassmann elements, i.e. $g^{(i)}=(g^{(i)})^*$ and $g^{(i)}a=ag^{(i)}$ for any $a\in \mathcal{A}_F$.

Now recall the adjoint map Ad: $\mathcal{U}(\mathcal{A}_F)\ni u \mapsto U_F\equiv\pi(u)J_F\pi(u)J_F^*\in \text{End}(\mathcal{H}_F)$, and note that for brevity the representation symbol $\pi$ will be implicit when no danger of confusion is present.
Then, for $h=\begin{pmatrix}
h_1 \\
h_2
\end{pmatrix} \in \mathcal{H}_F$ it is readily checked that:
\begin{itemize}
\item $U_F^*U_F=U_FU_F^*=\hbox{\boldmath $1$}_{ F}$,
\item $U_F^*\gamma U_F=\gamma$,
\item $U_F^*J_F U_F=J_F$,
\item $U_F h=\begin{pmatrix}
u_1h_1u_2^* \\
u_2h_2u_1^*
\end{pmatrix}$.
\end{itemize}
Given the last property above, computing $\text{Ker}(\text{Ad}) = \lbrace u\in\mathcal{U}(A_F) : U_F h=h \quad \text{for all} \quad h\in H_F \rbrace$ yields the conditions $g^{(1)}=g^{(2)}\equiv g_e$, i.e. an element of the kernel has the form,
\[
\text{Ker}(\text{Ad})\in \begin{pmatrix}
\e^{ig_e} &0 \\
0& \e^{ig_e}
\end{pmatrix}.
\]
Now, the gauge group of $\mathcal{A}$ is defined to be
\[
\mathcal{G}(\mathcal{M}^{3|2}\otimes \mathcal{F}_F) \equiv \{U=uJuJ^* | u\in \mathcal{U}(\mathcal{A}) \},
\]
but $\mathcal{G}(\mathcal{A}_M)$ is trivial, and since $\text{Ker}(\text{Ad})=\mathcal{U}(\mathcal{A}_F)_{J_F}\equiv\{ u\in\mathcal{A}_F : uJ_F=J_Fu^* \}$, the gauge group of the finite space is given by
\[
\mathcal{G}(\mathcal{F}_F) = \mathcal{U}(\mathcal{A}_F) / \text{Ker}(\text{Ad}).
\]
It is immediate to calculate that an element $\underline{u} \in \mathcal{G}(\mathcal{F}_F)$ is of the form
\begin{eqnarray*}
\underline{u} =
 \begin{pmatrix}
e^{\frac{i}{2}g}&0 \\
0&e^{-\frac{i}{2}g}
\end{pmatrix},
\end{eqnarray*}
where $g\equiv g^{(1)}-g^{(2)}$,
and, if $\underline{U}=\underline{u}J_F\underline{u}J_F^*$, then
\[
\underline{U}h=\begin{pmatrix}
e^{ig}h_1 \\
\e^{-ig}h_2
\end{pmatrix}.
\]

It is interesting to note that the chiral restriction imposed on a superspinor is not consistent with gauge covariance. Indeed,
the compatibility condition
\[
\e^{ig}D_\alpha \tilde\Psi^\alpha =D_\alpha \e^{ig}\tilde\Psi^\alpha,
\]
is satisfied if and only if $D_\alpha g = 0.$ In the case of a real, even superfield $g$ the latter condition yields, after a short calculation, that $g$ has to be a real, constant element of $\Lambda_\infty.$

 \subsection{The fermionic action}
Since there is no concern with regards to the so-called fermion doubling problem which is encountered when one reproduces the standard model by the techniques of NCG, here the fermionic action is taken in it's original form:
\[
\left\langle \xi , \mathcal{D}_A \xi \right\rangle,
\]
for $\xi \in \mathcal{H}=\mathcal{H}_M\otimes \mathcal{H}_F$.
Such elements have the form
\begin{align}
\xi &= \Psi(x,\theta)\otimes h = \Psi_+\otimes e + \Psi_-\otimes \bar{e}, \\
	&= \begin{pmatrix}  \Psi\otimes h_1&0\\0&\Psi\otimes h_2	\end{pmatrix} = \begin{pmatrix} \Psi_+&0 \\ 0&\Psi_-	\end{pmatrix},
\end{align}
where $\{ e,\bar{e} \}$ is an orthonormal basis for $\mathcal{H}_F$, such that $e\in \mathcal{H}_F^+$ and $\bar{e}\in \mathcal{H}_F^-$, (i.e. $\gamma_F e = e$ and $\gamma_F \bar{e}=-\bar{e}$), and such that $J_F e=\bar{e}$ and $J_F \bar{e} = e$.  Also, recall that each $\Psi_\pm \in \mathcal{H}_M$ is a super-spinor with the form
\[
\Lambda^o_\infty \ni \Psi^\alpha_\pm(x,\theta)=\psi^\alpha_\pm(x)+F^\alpha_{\pm\:\beta}(x)\theta^\beta+\chi^\alpha_\pm(x)\theta\theta,
\]
which means that $\psi^\alpha_\pm(x),\chi^\alpha_\pm(x)\in \Lambda_\infty^o$ and $F^\alpha_{\pm\:\beta} (x) \in \Lambda_\infty^e$.

Given the fluctuated Dirac operator computed previously,
\begin{align*}
\mathcal{D}_A &=i\gamma_m \begin{pmatrix}
\partial_m & 0 \\ 0 & \partial_m
\end{pmatrix}+\begin{pmatrix}
\gamma^m A_m & 0 \\
0 & -\gamma^m A_m
\end{pmatrix} \\
&=\mathcal{D}+\gamma^m A_m \otimes \gamma_F \quad \text{where} \quad A_m =i(a_1\partial_m b_1 - a_2\partial_m b_2),
\end{align*}
the fermionic action is calculated to be
\[
\left\langle \xi , \mathcal{D}_A \xi \right\rangle = \left\langle \xi,(\mathcal{D}_M\otimes \hbox{\boldmath $1$}_{ F})\xi \right\rangle + \left\langle \xi, (\gamma^m A_m \otimes \gamma_F)\xi \right\rangle.
\]
For the first term,
\begin{align*}
\left\langle \xi,(\mathcal{D}_M \otimes \hbox{\boldmath $1$}_{ F}) \xi \right\rangle &=  \left\langle \Psi_+ ,\mathcal{D}_M
\Psi_+ \right\rangle \left\langle  e, e \right\rangle + \left\langle \Psi_- ,\mathcal{D}_M
\Psi_- \right\rangle \left\langle  \bar{e}, \bar{e} \right\rangle \\
&=\left\langle \Psi_+ ,\mathcal{D}_M
\Psi_+ \right\rangle + \left\langle \Psi_- ,\mathcal{D}_M \Psi_- \right\rangle \\
&\equiv \left\langle \Psi_\pm ,\mathcal{D}_M
\Psi_\pm \right\rangle,
\end{align*}
and similarly the second term,
\begin{align*}
\left\langle \xi, (\gamma^m A_m \otimes \gamma_F)\xi \right\rangle &= \left\langle \Psi_+ , \gamma^m A_m\Psi_+ \right\rangle  + \left\langle \Psi_- , \gamma^m A_m \Psi_- \right\rangle \\
&\equiv \left\langle \Psi_\pm , \gamma^m A_m\Psi_\pm \right\rangle.
\end{align*}

As before, SUSY invariance of the action under a supersymmetry transformation is guaranteed for terms which are of highest order in the Grassmann variables.  But since (for the KO-dim 6 case) $A_m=i(a_1\partial_m b_1-a_2\partial_m b_2)$ where $a_i,b_i\in\Lambda_\infty^e$ for $i=1,2$, $A_m$ is itself represented by an even superfield on $\mathbb{R}^{3|2}$ and can be written in the form
\[
A_m=\mathsf{A}_m+\lambda_{m,\alpha}\theta^\alpha+\mathsf{B}_m\theta\theta,
\]
for some independent fields $\mathsf{A}_m$, $\lambda_{m,\alpha}$, and $\mathsf{B}_m$.

In these terms
\begin{align*}
\left\langle \xi, (\gamma^m A_m \otimes \gamma_F)\xi \right\rangle_{\theta\theta} &= \left\langle \psi_\pm , \gamma^m \mathsf{B}_m\psi_\pm \right\rangle + \left\langle \psi_\pm, \gamma^m \lambda_{m,[2}F_{\pm\:1]} \right\rangle +\left\langle \psi_\pm , \gamma^m \mathsf{A}_m\chi_\pm \right\rangle  \\
&+ \left\langle F_{\pm\:[1},\gamma^m\lambda_{m,2]}\psi_\pm \right\rangle + \left\langle F_{\pm\:[2},\gamma^m \textsf{A}_m F_{\pm\:1]} \right\rangle + \left\langle \chi_\pm,\gamma^m \mathsf{A}_m\psi_\pm \right\rangle.
\end{align*}

And finally, we may write down the complete SUSY invariant fermionic action
\begin{align*}
\left\langle\xi,\mathcal{D}_A\xi\right\rangle_{\theta\theta}&= \left\langle \psi_\pm ,\mathcal{D}_M\chi_\pm \right\rangle + \left\langle  F_{\pm\: [2} , \mathcal{D}_M F_{\pm\: 1]} \right\rangle + \left\langle \chi_\pm,\mathcal{D}_M\psi_\pm \right\rangle  \\
&+ \left\langle \psi_\pm , \gamma^m \mathsf{B}_m\psi_\pm \right\rangle + \left\langle \psi_\pm, \gamma^m \lambda_{m,[2}F_{\pm\:1]} \right\rangle +\left\langle \psi_\pm , \gamma^m \mathsf{A}_m\chi_\pm \right\rangle  \\
&+ \left\langle F_{\pm\:[1},\gamma^m\lambda_{m,2]}\psi_\pm \right\rangle + \left\langle F_{\pm\:[2},\gamma^m \textsf{A}_m F_{\pm\:1]} \right\rangle + \left\langle \chi_\pm,\gamma^m \mathsf{A}_m\psi_\pm \right\rangle.
\end{align*}

  \subsection{The spectral action}
Within the Connes approach the dynamics of a gauge field is encoded in the \textit{spectral action},
\[
S_A = {\rm Tr}f(\mathcal{D}_A),
\]
where $f$ is a smooth, rapidly vanishing function whose moments determine the parameters (e.g.\ coupling constants) of the discussed model.
In the present situation of a flat (super)space the calculation of $S_A$ essentially trivializes, and boils down to calculating the trace of 
the third power of the fluctuated Dirac operator. 
Since
\begin{eqnarray*}
(D_A)^3
& = &
\frac{i}{2}\left(-\partial^2\partial_m+(A\cdot A)\partial_m  +2 A_m (A\cdot \partial) + \partial_m(A\cdot A) + A_m (\partial A) \right)  \gamma^m\otimes\hbox{\boldmath $1$}_{ F}
\\[4pt]
& - &
\frac12 \left(2A_m\partial^2+2(A\cdot\partial)\partial_m +(\partial A)\partial_m  +2 \partial_m(A\cdot\partial)+\partial_m(\partial A)- 2(A\cdot A)A_m\right)  \gamma^m\otimes\gamma_{ F}
\\[4pt]
& - &
\frac12 \partial_p F_{mn}\left(\gamma^p\left[\gamma^m,\gamma^n\right]\right)\otimes \gamma_{ F}
+
\frac{i}{2} A_p F_{mn}\left(\gamma^p\left[\gamma^m,\gamma^n\right]\right)\otimes \hbox{\boldmath $1$}_{ F},
\end{eqnarray*}
where $F_{mn} = \partial_m A_n - \partial_n A_m,$
we get
\[
S_A \sim \int\,\epsilon^{pmn}A_p F_{mn} d^3x.
\]
The SUSY invariant action in the gauge sector is thus of the form
\begin{eqnarray*}
\left(S_A\right)_{\theta\theta} 
& = &
\frac{1}{g} \int \epsilon^{pmn}\left(
\mathsf{A}_p(\partial_m\mathsf{B}_n - \partial_n\mathsf{B}_m) 
+ 
\mathsf{B}_p(\partial_m \mathsf{A}_n - \partial_n\mathsf{A}_m)
-
\lambda_{p,\alpha}(\partial_m \lambda_n^\alpha - \partial_n \lambda_m^\alpha)
\right),
\end{eqnarray*}
where $g$ is a coupling constant. This action is again automatically invariant under the SUSY transformation of $ \mathsf{A}_m, \lambda_{m,\alpha}$ and $\mathsf{B}_m$ defined by
\[
\delta A_m(x,\theta) = \delta\mathsf{A}_m+\delta\lambda_{m,\alpha}\theta^\alpha+\delta\mathsf{B}_m\theta\theta = A_m(x+\delta x,\theta + \delta\theta) - A_m(x,\theta),
\]
where $\delta x$ and $\delta\theta$ are of the form (\ref{SUSY:on:coordinates}).

\section{Conclusions}

The preceding work proposes and exemplifies a strategy for the incorporation of a superspace formulation of the principle of supersymmetry into the formalism of noncommutative geometry, up to and including the spectral action.  This has been done in as simple a setting as possible, not solely for computational convenience, but as well, so as to avoid obfuscation of the guiding principles and machinery of the noncommutative method.

In fact, the perspicacious reader will have undoubtedly (and rightly) noted that there is nothing truly noncommutative in the example which is presently investigated.  Through consideration of a less trivial finite space (e.g. \textit{Supermatrix} algebras), one may introduce noncommutativity into the picture and expect the resulting theory to have a richer structure (e.g. non-abelian gauge fields and a Higgs sector analogue).

Similarly, the Dirac operator being considered here is a somehow naive choice as it only contains derivatives over the commuting coordinates.  A consequence of this choice, it should be emphasized, is that the resulting field theory is nothing like a physically relevant one.  For example, electrodynamics is the usual result of the AC-geometry approach when a 2-point finite space is considered. But by choosing the spinorial derivative as Dirac operator, absent are terms of the form $\psi^\alpha D \psi^\alpha$. We expect that an honest superspace Dirac operator built from supercovariant derivatives will further extend the richness, and, it is hoped the physical relevance, of any theory developed according to this strategy.  Investigation of such an operator is in progress and results will be presented in due course.

\section*{Acknowledgement}
The author would like to thank L. Hadasz for his assistance in the preparation of this work.

\end{document}